\documentclass[%
reprint,
showpacs,preprintnumbers,
amsmath,amssymb,
aps,
]{revtex4-1}
\usepackage{graphicx}
\usepackage{dcolumn}
\usepackage{bm}
\usepackage{color}
\usepackage{ulem}

\begin{document}
\preprint{APS/123-QED}

\title{Argument on superconductivity pairing mechanism from cobalt impurity doping in FeSe: spin ($s_{\pm}$) or orbital ($s_{++}$) fluctuation}

\author{T. Urata$^1$}

\author{Y. Tanabe$^1$}
 \thanks{Corresponding author: ytanabe@m.tohoku.ac.jp}

\author{K. K. Huynh$^2$}

\author{Y. Yamakawa$^3$}

\author{H. Kontani$^3$}

\author{K. Tanigaki$^{1, 2}$}
\thanks{Corresponding author: tanigaki@sspns.phys.tohoku.ac.jp}

\affiliation{$^1$Department of Physics, Graduate School of Science, Tohoku University, Sendai, 980-8578, Japan}

\affiliation{$^2$WPI-Advanced Institutes of Materials Research, Tohoku University, Sendai, 980-8577, Japan}

\affiliation{$^3$Department of Physics, Nagoya University, Furo-cho, Nagoya 464-8602, Japan}

\date{\today}
 
\begin{abstract}
In high-superconducting transition temperature ($T_{\rm c}$) iron-based superconductors, interband sign reversal ($s_{\rm \pm}$) and sign preserving ($s_{\rm ++}$) $s$-wave superconducting states have been primarily discussed as the plausible superconducting mechanism.
We study Co impurity scattering effects on the superconductivity in order to achieve an important clue on the pairing mechanism using single crystal Fe$_{1-x}$Co$_x$Se and depict a phase diagram of a FeSe system.
Both superconductivity and structural transition / orbital order are suppressed by the Co replacement on the Fe sites and disappear above $x$ = 0.036.
These correlated suppressions represent a common background physics behind these physical phenomena in the multiband Fermi surfaces of FeSe.
By comparing experimental data and theories so far proposed, the suppression of $T_{\rm c}$ against the residual resistivity is shown to be much weaker than that predicted in the case of a general sign reversal and a full gap $s_{\pm}$ models.
The origin of the superconducting paring in FeSe is discussed in terms of its multiband electronic structure.
\end{abstract} 

\pacs{}
\maketitle

\section{Introduction}
The superconducting paring mechanism of high temperature superconductivity has been a long-lasting priority research area, and is one of the most important and intriguing scientific subjects.
After discovery of Fe-based superconductors (FeSCs), the superconducting mechanism has been discussed from the viewpoint of their unique multiband Fermi surfaces.
The superconducting gap functions primarily discussed in FeSCs are the interband sign reversal $s$-wave ($s_{\rm \pm}$) and the sign preserving $s$-wave ($s_{\rm ++}$) states. \cite{Mazin,Kuroki,Kontani}.
A stripe type antiferromagnetic spin fluctuation has been considered to mediate $s_{\pm}$-wave state, while Fe 3$d$ orbital fluctuation mediates the $s_{++}$-wave state.

Impurity scattering in superconductivity gives important information for understanding the pairing mechanism.
A phonon mediated isotropic ``BCS'' superconductivity is robust against the nonmagnetic impurity scattering.
Meanwhile, since the Cooper pair is glued by the $k$-dependent anisotropic scattering in the sign reversal superconductivity, e.g. ($\pi, \pi$) spin fluctuation in cuprates, similar impurities induce random scattering by ending up with a strong pair breaking \cite{Ohkawa,Ishida_imp}.
A general theory of pair breaking in the latter case was given by Abrikosov and Gorkov (AG-theory) \cite{AG}.
In the FeSCs, impurities are expected to induce a significant reduction in superconducting transition temperature ($T_{\rm c}$) for the $s_{\pm}$-wave states.
Although the nonmagnetic impurity doping effects have been mostly examined by the AG-theory in connection to the scaling relation between $T_{\rm c}$ and residual resistivity ($\rho_0$) \cite{M_Sato,Y_Nakajima,Kirshenbaum,Inabe}, the important intrinsic multiband nature has been frequently neglected.

\begin{figure*}[t]
\includegraphics[width=1.0\linewidth]{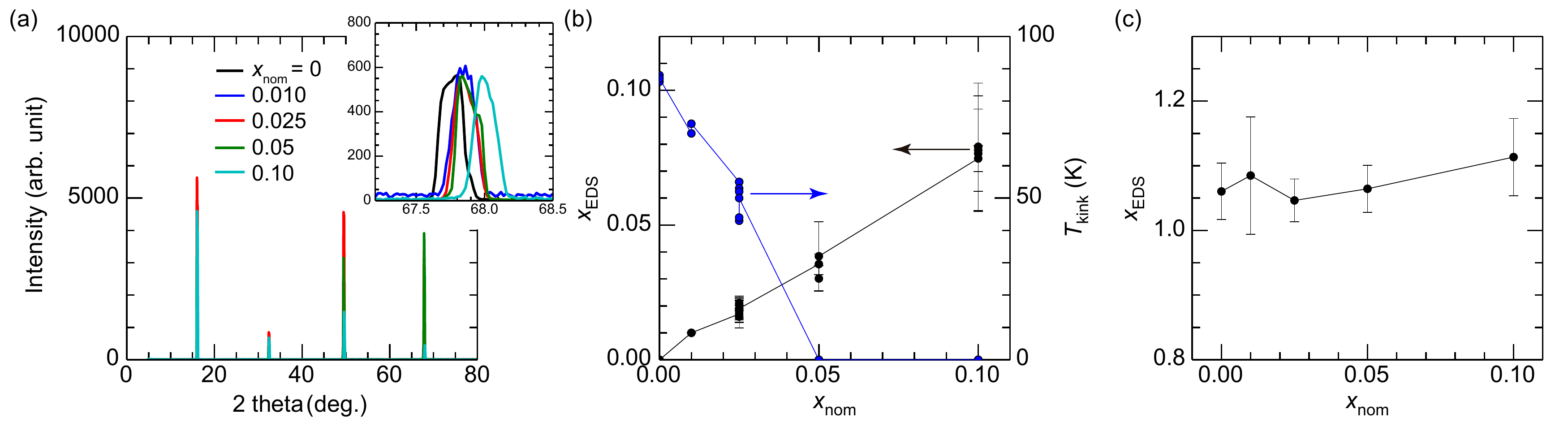}
\caption{
(a) The X-ray diffraction data of Fe$_{1-x}$Co$_x$Se ($x_{\rm nom}$ = 0, 0.010, 0.025, 0.05, and 0.10) around the c-axis. The inset represents the normalized (004) peaks.
(b) The nominal concentration ($x_{\rm nom}$) dependence determined by EDS ($x_{\rm EDS}$) and $T_{\rm kink}$.
$T_{\rm kink}$ is defined as the temperature showing a kink in the resistivity curve as described in the later section.
(c) The nominal concentration dependence of concentration Fe+Co with respect to Se (i.e. (Fe+Co)/Se).
The error bars are defined as the standard deviations of measurements.}
\label{cha}
\end{figure*}

FeSe is one of the FeSC families showing superconductivity at around 9 K \cite{Hsu_FeSe}.
In the vicinity of the tetragonal / orthorhombic structural transition temperature ($T_{\rm s} \approx 90$ K), the orbital order has been recognized to develop without long-range antiferromagnetic (AFM) order, being in strong contrast with the other FeSCs \cite{Kotegawa,McQueen,Nakayama,Shimojima_FeSe,Baek,Bohmer_NMR}.
A sign preserved superconducting state has been indicated by scanning tunneling microscopy (STM) and spectroscopy (STS) in single layer FeSe \cite{Q_Fan}.
Nevertheless, in thicker films and bulk FeSe, the nodal superconducting gap has been suggested by STM/STS as well as the London penetration depth, implying the contribution of AFM spin fluctuation to the formation of the Cooper pairs \cite{Song,Kasahara_PNAS}.
Both neutron scattering and nuclear magnetic resonance have shown AFM spin fluctuations \cite{Rahn,Wang,Baek,Bohmer_NMR}, whereas an imperfect nesting between electron and hole Fermi surfaces has been observed by orbital resolved angle resolved photo emission spectroscopy (ARPES) \cite{Suzuki}.
Detailed studies on Co doping are considered to give an important hint for understanding the mechanism of superconductivity presently debated in FeSe.

In the present paper, we report systematic electrical transport studies on the effect of Co impurity doping for single crystal FeSe.
In the FeAs system, Co has been regarded as a nonmagnetic impurity and provides an additional electron as a carrier \cite{Sekiba}, being in contrast with the situation of Mn \cite{Texier,Urata_kondo}.
Present Hall coefficients and ARPES \cite{R_Peng,Phan_JPS} support this understanding.
Therefore, Co is considered to act as a nonmagnetic impurity and add an electron as an itinerant carrier to FeSe.
Our present experimental data indicate a correlation between the suppression of $T_{\rm c}$ and that of the structural transition orbital order when Fe is replaced by Co.
By carefully analyzing the dependencies between the suppression of $T_{\rm c}$ and $\rho_0$ and taking into consideration the realistic electron and hole Fermi surfaces in FeSe \cite{Nakayama_unp}, we found that the suppression of $T_{\rm c}$ against $\rho_0$ is clearly weaker than those expected from both the AG-theory and a recent more particular model for the $s_{\rm \pm}$-wave state \cite{Yamakawa}.
Our experimental observations give better agreement with the sign preserved superconducting gap states and suggest important multiorbital nature of Fermi surfaces for the occurrence in superconductivity of FeSe \cite{Kontani,Fernandes,Onari}.

\section{Experiments}
High quality single crystals of Fe$_{1-x}$Co$_x$Se ($0 \leq x \leq 0.075$) were grown by a molten salt flux method \cite{Bohmer_synth,Huynh,Nakayama}.
The temperature of hot and cold positions was kept for 390 and 240$^\circ$C, respectively.
After $\sim$ 10 days, single crystals were grown around the cold part of the quartz tube.
Being different from the conventional method, polycrystalline samples \cite{Mizuguchi} were employed as a precursor.
Note that $x_{\rm nom}$ indicates the nominal composition of Co applied for synthesis of polycrystals.
We examined the quality of the prepared samples by X-ray diffraction (XRD, Cu K-$\alpha$ radiation wavelength of 1.5406 \AA) around the $c$-axis as well as energy dispersive X-ray spectroscopy (EDS).
The magnetic susceptibility was measured at $B$ = 1T parallel to the ab-plane.
The temperature dependence of electrical resistivity and Hall resistance was measured by a four probe method.
The superconducting transition temperatures ($T_{\rm c}$s) were determined at the end point of the superconducting transition with a value of approximately less than 1.0$\times 10^{-8}$ $\Omega$cm.
In the Hall resistance measurements, magnetic fields were varied in a range of $|B|\leq 9$ T parallel to the c-axis.


\section{Results}
\subsection{Sample Characterization}
Figure \ref{cha} (a) shows XRD spectra obtained for $x_{\rm nom}$ = 0, 0.01, 0.025, 0.05, and 0.10.
Clear (00$l$) ($l$ = 1-4) peaks were observed.
No impurity peaks were detected within the experimental errors.
Since the lattice shrinkage is very small in the case of Co doping \cite{Mizuguchi}, it was difficult to see a significant influence on their lattice parameters among samples with small concentration of Co even though the shift in the (004) peak between $x_{\rm nom}$ = 0 and 0.10 was detected.
The black squares in Fig. \ref{cha} (b) represent the dependence of $x_{\rm nom}$ as a function of Co concentration determined by EDS ($x_{\rm EDS}$).
EDS spectra were taken at 10 different positions on each specimen.
The error bars were estimated as the standard deviations from the average.
A monotonic increase in $x_{\rm EDS}$ was observed against $x_{\rm nom}$.
Note that the nominal concentration of $x_{\rm nom}$ = 0.01 is used for the sample instead of $x_{\rm EDS}$ because EDS was not sufficiently sensitive for detecting the Co peak under such a small concentration.
The blue circles in Fig. \ref{cha} (b) show that the $T_{\rm kink}$ (the kink temperature in the resistivity curve as shown in the section of the electrical resistivity) reduced monotonically with $x_{\rm nom}$.
These experimental data indicate that the systematic substitution of Co was successfully accomplished.
We regard $x_{\rm EDS}$ as the real $x$ and used this for discussion hereafter.
In the parent FeSe grown by a KCl/AlCl$_3$ flux method, no interstitial Fe was found by X-ray diffraction measurements \cite{Bohmer_synth}.
In order to check the interstitial Fe carefully in Fe$_{1-x}$Co$_x$Se, the concentration ratio: (Fe+Co)/Se was compared against the nominal Co concentrations as shown in Fig. \ref{cha} (c).
Since the ratio is almost constant within the error bars for all nominal concentrations, the interstitial Fe is not influenced by the Co doping.
It should be emphasized that we selected the samples with (Fe+Co)/Se $< 1.05$ for electrical transport measurements.

\subsection{Electrical Resistivity}
\begin{figure}[h]
\includegraphics[width=1.0\linewidth]{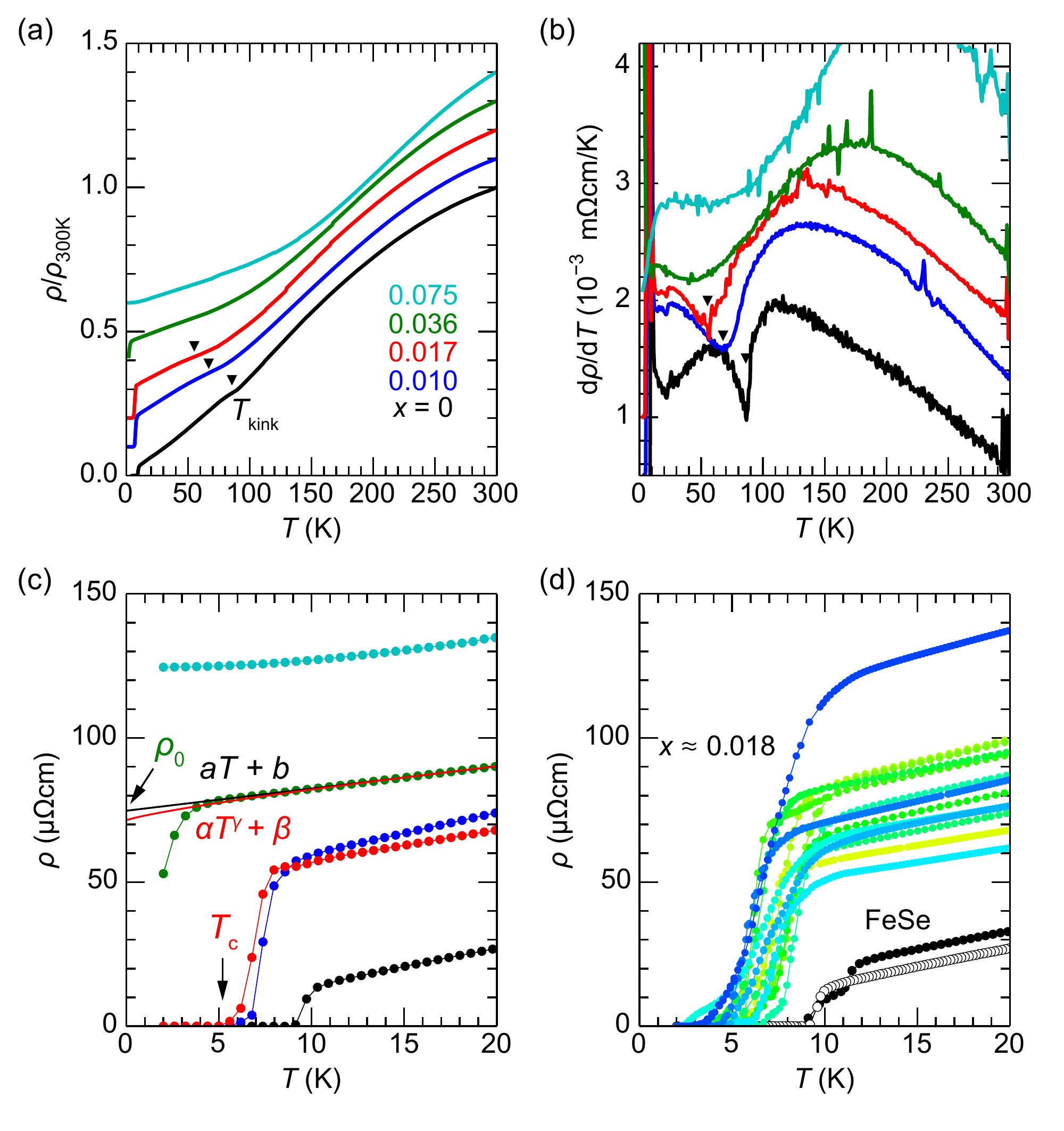}
\caption{(a) Normalized temperature dependence of resistivity ($\rho/\rho_{\rm 300K}$). Note that each curve is shifted by 0.1 for clarity. The closed triangles indicate the kink in the resistivity curve ($T_{\rm kink}$).
$T_{\rm kink}$ is defined as the peak position of the 1st $T$ derivative of $\rho$.
(b) Temperature dependence of the 1st derivative of $\rho$.
(c,d) Enlarged view of temperature dependence of $\rho$. The curve is not shifted. (c) The black and the red lines show fitting results of $\rho$-$T$ curves in the normal states using $\rho$ = ($aT+b$) or $\rho$ = ($\alpha T^\gamma + \beta$), respectively.
(d) Closed and open black circles represent $x$ = 0. The others (blue, yellow, and green) indicate $x_{\rm nom}$ = 0.025. Their averaged concentration was $x$ $\approx$ 0.018.
}
\label{res}
\end{figure}
Temperature dependence of the normalized electrical resistivity ($\rho/\rho_{\rm 300K}$) for Fe$_{1-x}$Co$_x$Se ($0 \leq x \leq 0.075$) is shown in Fig. \ref{res}(a).
In parent FeSe, the $\rho/\rho_{\rm 300K}$ showed a kink ($T_{\rm kink}$) due to the structural transition \cite{Hsu_FeSe,Bohmer_synth}.
This can clearly be seen as a dip of the first derivative as shown in Fig. \ref{res}(b).
$T_{\rm kink}$ decreased with an increase in Co concentration and finally disappeared at $x$ = 0.036, indicating that the high-temperature tetragonal phase holds above $x$ = 0.036.
As seen in the magnified $\rho$-$T$ plots (Fig. \ref{res}(c)) at low temperatures, the $T_{\rm c}$ decreases with an increase in Co and the superconductivity is killed at $x$ = 0.036 above 2 K.
Both suppression of superconductivity and structural transition in FeSe occur by partial replacement of Fe to Co.
In order to interpret the $T_{\rm c}$ suppression by Co in terms of the impurity scattering, residual resistivity $\rho_0$ was evaluated.
The absolute values of resistivity ($\rho$) as a function of temperature are very sensitive to the measurement conditions and generally contain large errors arising from sample shape or/and inhomogeneity, etc..
In order to estimate a reliable value, resistivity measurements were carefully performed on more than twenty samples selected from the same sample batch of $x_{\rm nom}$ = 0.025.
Figure \ref{res}(d) shows the results by excluding the samples showing large deviation from the average.
Just above the $T_{\rm c}$s, $\rho$ of $x_{\rm nom}$ = 0.025 ($x \approx 0.018$) are clearly larger than those of FeSe.

\begin{figure}[h]
\includegraphics[width=0.9\linewidth]{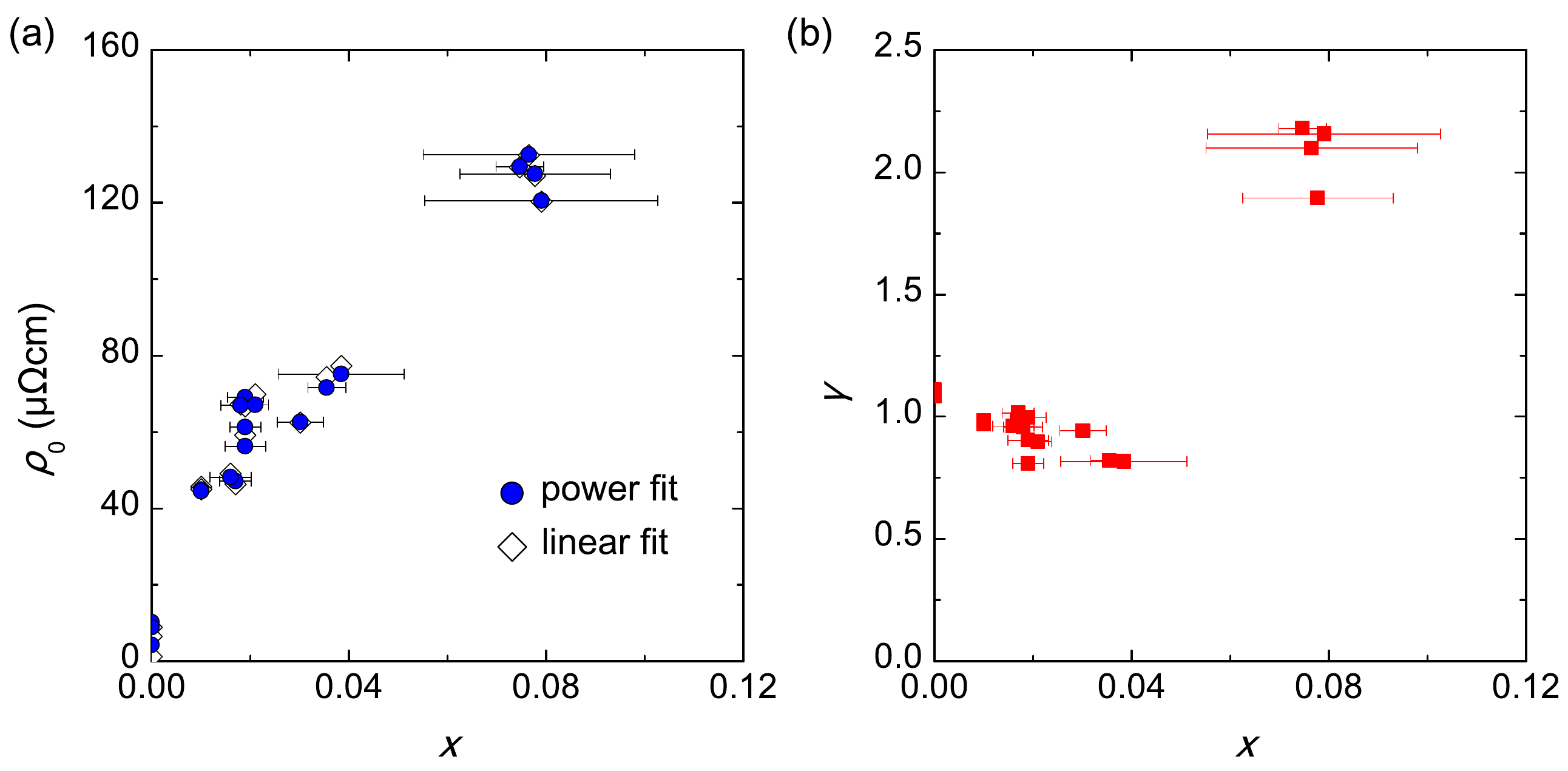}
\caption{(a) Co concentration ($x$) dependence of the residual resistivity, i.e. $b$ in the $aT+b$ and $\beta$ in the $\alpha T^{\gamma}+\beta$, versus $x$.
(b) The exponent $\gamma$ versus $x$.}
\label{rho0_eval}
\end{figure}

In order to deduce residual resistivity ($\rho_0$), two types of fitting were employed.
One is linear fitting using a function of $aT+b$, where $b$ represents $\rho_0$.
The other is fitting using a power function of $\alpha T^{\gamma}+\beta$, where $\beta$ represents $\rho_0$.
The fitting was made in the temperature range of 12$\sim$13K (just above the superconducting onsets) to 25$\sim$27 K.
An example of fitting line is shown in Fig. \ref{res}(c).
Figure \ref{rho0_eval} shows the $x$ dependence of these fitting parameters.
Although some errors can be recognized, $\rho_0$ apparently increases monotonically with $x$ as seen in Fig. \ref{rho0_eval}(a).
The value of $\rho_0$ was not significantly affected by the fitting functions.
The exponent $\gamma$ in the power function is depicted against $x$ in Fig. \ref{rho0_eval} (b).
At small concentrations it slightly decreases from $\sim$1.1 to $\sim$0.8 with $x$, and then jumped to $\gamma \approx 2.0$ at $x \approx 0.08$.

%

\subsection{Magnetic Susceptibility}
\begin{figure}[t] 
\includegraphics[width=0.8\linewidth]{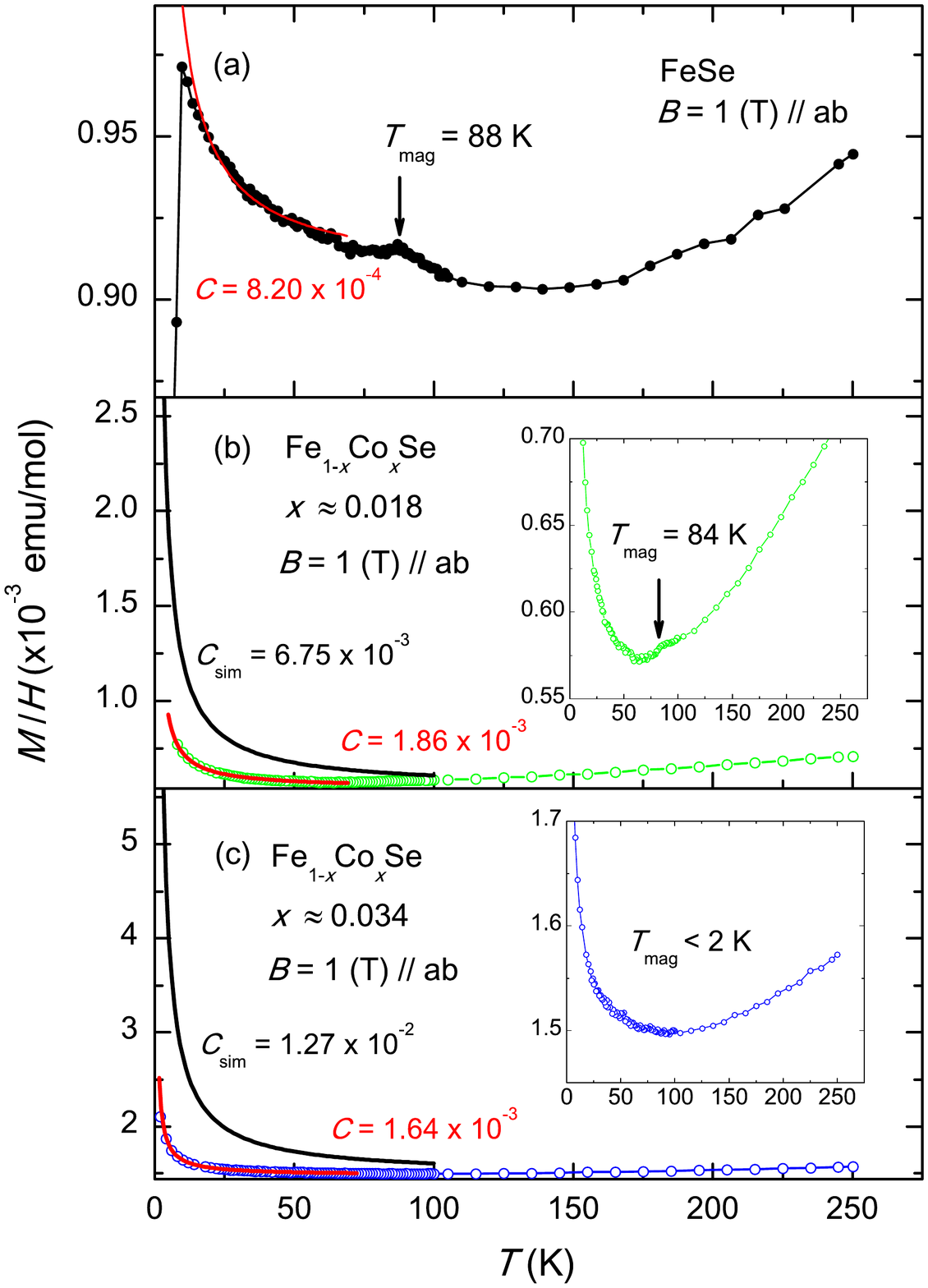}
\caption{Temperature dependence of magnetic susceptibility ($M/H$) for Fe$_{1-x}$Co$_x$Se with (a-c) $x\approx$ 0, 0.018, 0.034.
(b,c) Enlarged views are inserted.
$B$ = 1 T was applied in parallel to the $ab$-plane of crystals.
All data were taken on field cooling.
The $T_{\rm mag}$ is defined as the kink of the curves.
The red lines indicate Curie-Weiss fitting at low temperature, i.e. $C/T+A$ where $C$ and $A$ are the Curie constant and a certain real valued constant, respectively.
The black lines also represent the Curie-Weiss curves with $C_{\rm sim}$ simulated by assuming that Co is a magnetic impurity with the angular momentum quantum number $J$ = 1/2.
}
\label{magne}
\end{figure}

Temperature dependence of magnetic susceptibility ($M/H$) was measured on field cooling for Fe$_{1-x}$Co$_x$Se ($x\approx$ 0, 0.018, 0.034) as shown in Fig. \ref{magne}.
A kink was observed in the temperature dependence of $M/H$ curve at the structural transition temperature in FeSe, being in consistent with the earlier study \cite{Bohmer_synth}.
We express this temperature as $T_{\rm mag}$ in the present paper.
The $T_{\rm mag}$, the transition temperature in spin magnetic moments, was also found in our samples of $x\approx$ 0, 0.018, but it disappears in the sample of $x\approx$ 0.034.
This disappearance, most presumably corresponding to the absence of the tetragonal to orthorhombic structural transition, coincides with the disappearance of $T_{\rm kink}$ and $T^*$.

In order to confirm the mangetic statement of Co, $M/H$ curves were fitted in a low temperature range (15 -60 K) by a function $C/T + A$, where the Curie constant is assumed to be $C$ and $A$ is a certain constant value.
As shown in Fig. \ref{magne}, the fitting curves are represented as red lines and the obtained $C$ merely changes as a small value with an increase in $x$.
For comparison, the Curie constants were simulated ($C_{\rm sim}$) by assuming that all Co elements are localized with spin 1/2.
The evaluated black curves show large deviations from the experimental plots.
The obtained values of Curie constant were much smaller than the doped cobalt concentration and did not change significantly.
This result supports the scenario that the cobalt is acting as a nonmagnetic impurity in the Fe$_{1-x}$Co$_x$Se system.

\subsection{Hall Resistivity}

\begin{figure*}[t]
\includegraphics[width=1.0\linewidth]{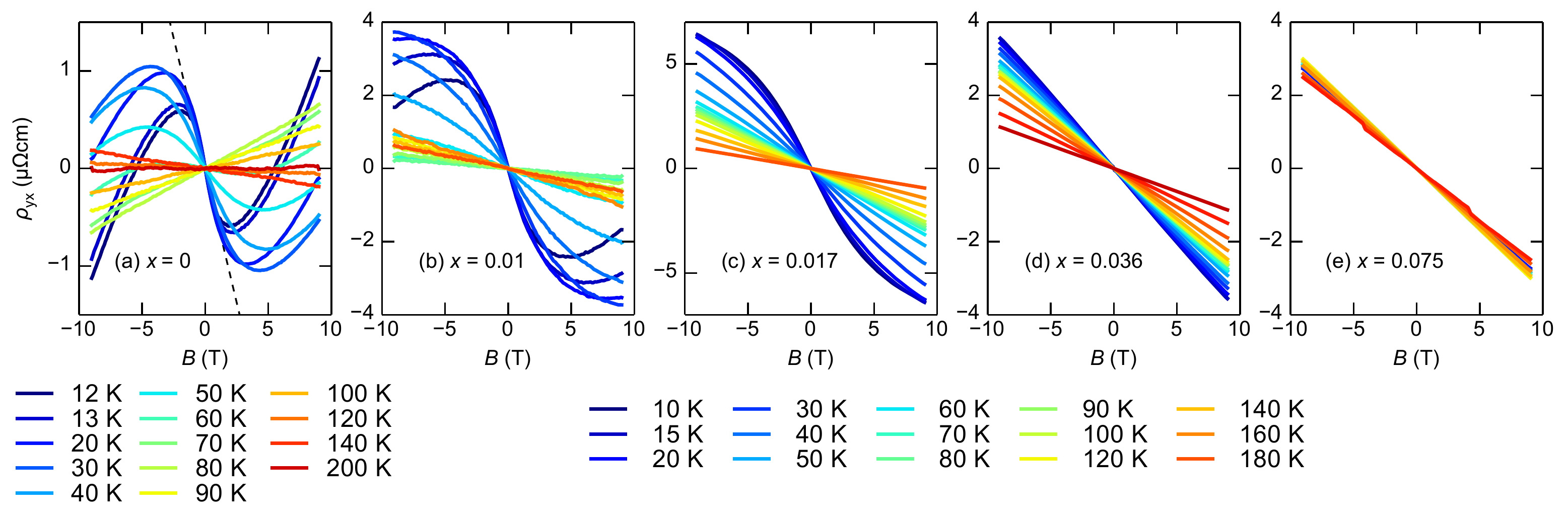}
\caption{Magnetic field ($B$) dependence of Hall resistivity ($\rho_{yx}$) of $x$ = 0, 0.010, 0.017, 0.036, and 0.075 ((a)-(e), respectively). 
The linear fitting line to derive the Hall coefficient ($R_{\rm H}$) in the low $B$ limit is shown as a dashed line in (a).
The $\rho_{yx}$s were measured at different temperatures for $x$ = 0 as noted in the legend at bottom left.
}
\label{Hall}
\end{figure*}

The magnetic field ($B$) dependence of Hall resistivity ($\rho_{yx}$) is shown in Fig. \ref{Hall}.
In order to remove the extrinsic deviations from symmetric components caused by the misalignment of electrodes, corrections were made on the raw $\rho_{yx}$ curves:

\begin{equation}
    \rho_{yx}(+|B|)=\frac{\rho_{yx}^{\rm raw}(+|B|)-\rho_{yx}^{\rm raw}(-|B|)}{2}.
\end{equation}
A strong nonlinear behavior at low temperatures was moderated as $x$ increases.
This can be understood to be caused by the suppression of mobility.
Moreover, temperature dependence was also moderated, being consistent with Hall coefficient's ($R_{\rm H}$'s) $T$ dependence described later.
$R_{\rm H}$ was determined from the slope of $\rho_{yx}$ at low $B$ where $\rho_{yx}$s develop linearly.
The linear fitting is displayed as a dashed line in Fig. \ref{Hall} (a).


\begin{figure}[h]
\includegraphics[width=0.8\linewidth]{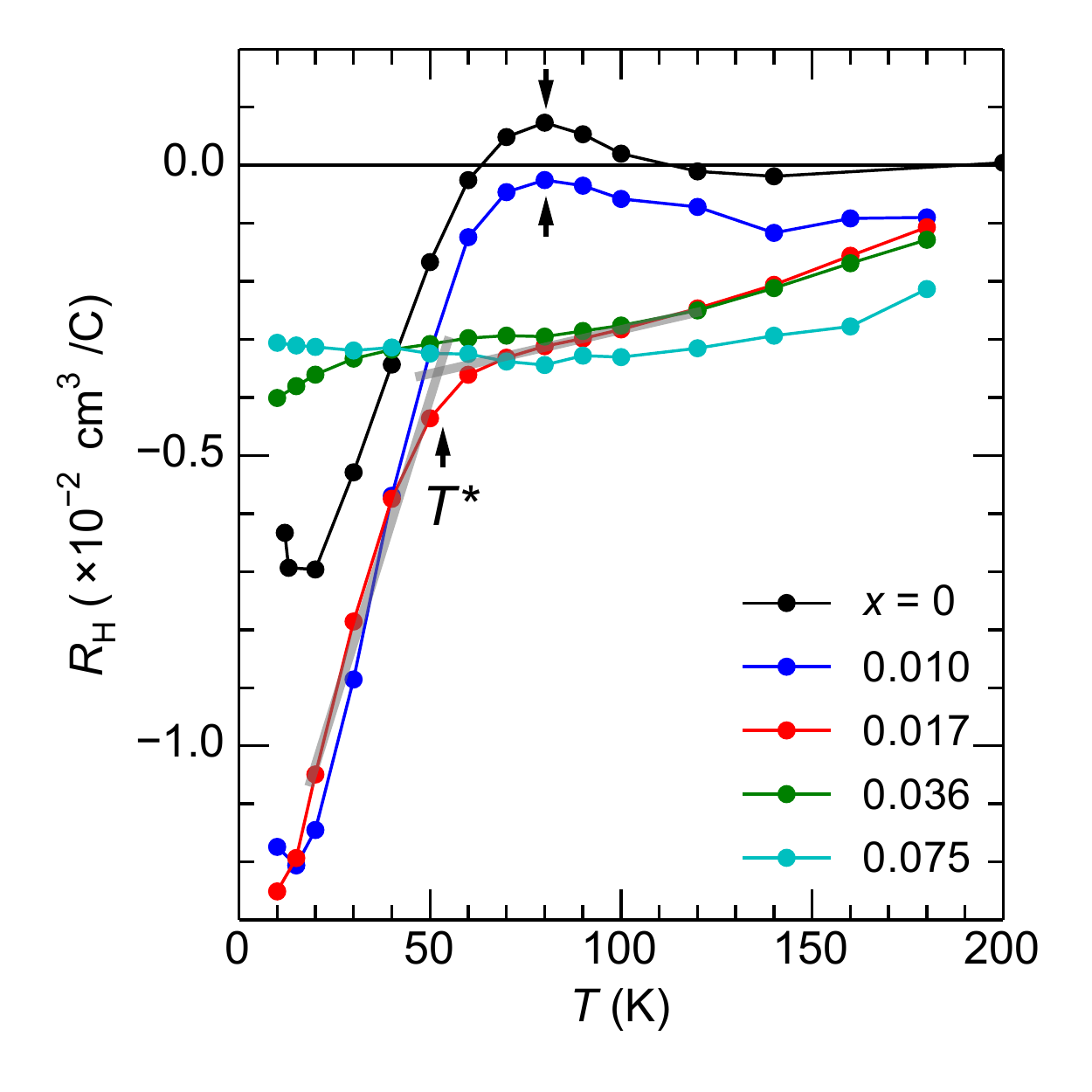}
\caption{Temperature dependence of Hall coefficient ($R_{\rm H}$) estimated in the low magnetic field regime.
The black arrows indite the temperature where $R_{\rm H}$ starts suddenly decreasing ($T^*$).
For $x$ = 0.017, $T^*$ is defined as the intersection point of two linear fitting lines shown as gray in color.}
\label{Hall_coeff}
\end{figure}

Temperature dependence of $R_{\rm H}$ is shown in Fig. \ref{Hall_coeff}.
For the parent FeSe, $R_{\rm H}$ became positive at around $T_{\rm kink}$ and dropped with decreasing temperature.
We define $T^*$ as the temperature where $R_{\rm H}$ shows a sudden decrease.
For $x$ = 0.017, $T^*$ is derived from the intersection of two linear fitting lines as shown in the gray lines of Fig. \ref{Hall_coeff}.
With an increase in Co concentration, $T^*$ lowered and almost disappeared at $x$ = 0.036, being consistent with the disappearance of $T_{\rm kink}$.
The amplitude of $R_{\rm H}$ at low temperatures once increased and then decreased with an increase in $x$. 
When the concentration of Co exceeded above $x$ = 0.036, the observed change in $R_{\rm H}$ is nearly suppressed.

\section{Analyses of $T_{\rm c}$ suppression rate by Co doping}
\begin{figure}[h]
\includegraphics[width=0.8\linewidth]{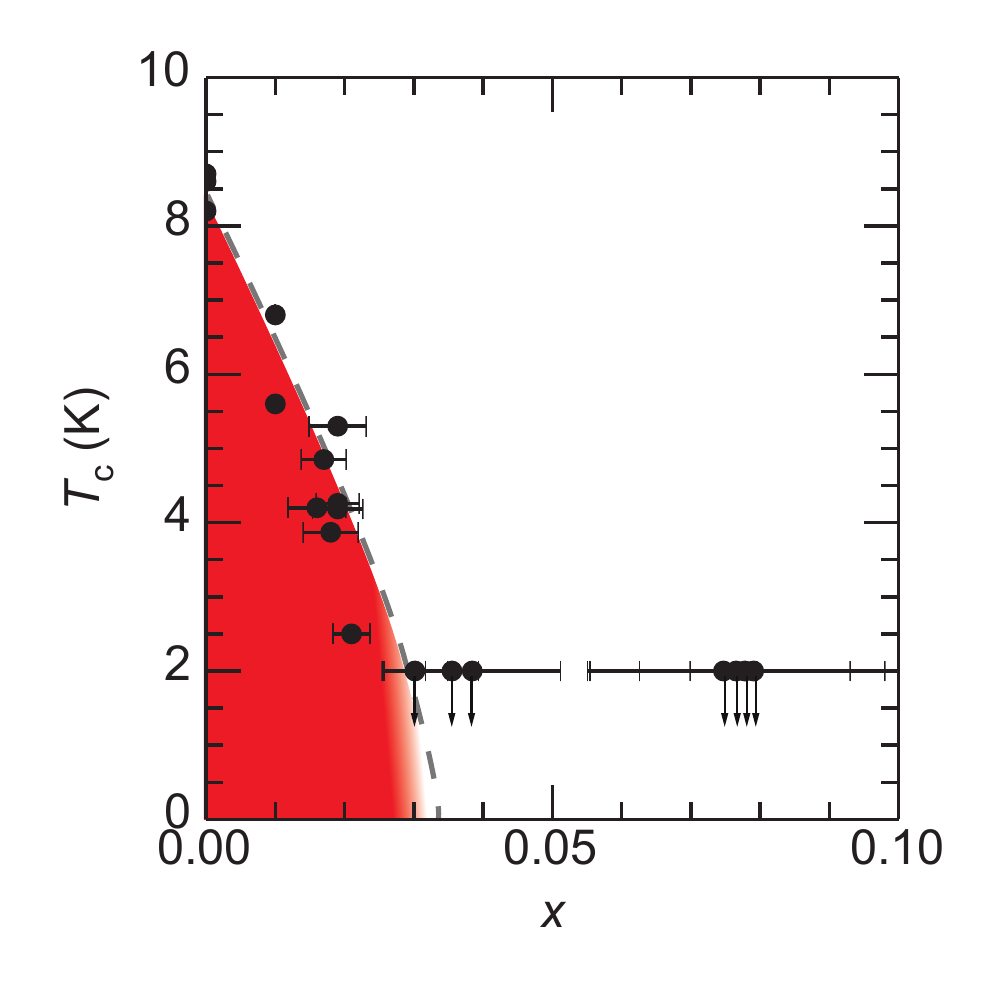}
\caption{
The Co concentration ($x$) dependence of the superconducting transition temperature ($T_{\rm c}$).
Down arrows indicate that the zero resistivity is not observed above 2 K.
The gray broken line is guide to the eyes.
}
\label{Tc_x}
\end{figure}

Figure \ref{Tc_x} shows the $x$ dependence of $T_{\rm c}$.
$T_{\rm c}$ monotonically decreased with an increase in $x$ within the experimental errors.
Figure \ref{rr} represents the $T_{\rm c}$ as a function of $\rho_{0}$.
The $\rho_{0}$ values were estimated from the intercepts of the linear fitting (white diamonds) and the $T^\gamma$ fitting (blue circles) on the resistivity curves at low temperatures as shown in Fig. \ref{res}(c).
$T_{\rm c}$ monotonically decreased with an increase in $\rho_{0}$ and disappeared at $\rho_{0}$ of $\approx$ 60 - 80 ${\rm\mu}\Omega$cm.

\subsection{AG-theory in realistic two band model}

In order to clarify the superconducting pairing mechanism in FeSe, here we evaluate how $T_{\rm c}$ is suppressed against residual resistivity when Co is doped as an impurity in the light of two models.
The first model is the AG-theory which can be applicable to the sign reversal superconducting states \cite{AG}.
The other is the specific model formulated for the full gap $s_{\pm}$-wave states theoretically calculated based on the realistic physical condition of FeSCs \cite{Yamakawa}.
Both models are valid in the limit that nonmagnetic impurities are not incorporated to a great extent.
The AG-theory is formulated as,

\begin{equation}
    {\rm ln}\left(\frac{T_{\rm c}^{\rm AG}}{T_{\rm c}^0}\right) = \Psi\left(\frac{1}{2}\right)-\Psi\left(\frac{1}{2} + \frac{\hbar}{4\pi\tau k_{\rm B}T_{\rm c}^{\rm AG}}\right),
\label{AG_eq}
\end{equation}
where $k_{\rm B}$ is the Boltzmann constant, $T_{\rm c}^0$ is the superconducting transition temperature in the pristine material, $\tau$ is the scattering time, and $\Psi(z)$ is the digamma function.
Here we use superconducting transition temperature as $T_{\rm c}^{\rm AG}$ to differentiate from experimental values ($T_{\rm c}$).
The pair breaking energy is characterized by $\tau$.
The value of $\tau$ was computed at each value of $T_{\rm c}^{\rm AG}$ using Eq. \ref{AG_eq} with $T_{\rm c0} = 9$ K.
In order to compare the AG-theory and our experimental data, $\tau$ needs to be converted into the residual resistivity in a reasonable way as follows.

In FeSe, nearly semimetallic electronic structure is realized \cite{Nakayama,Shimojima_FeSe,Huynh,Kasahara_PNAS,Terashima}.
The ARPES reported a single electron Fermi surface (FS) around the M-point and a single hole FS around $\Gamma$.
It should be mentioned that the recent mobility spectrum analysis and the quantum oscillation measurements in magnetoresistance reported an additional small FS \cite{Huynh,Watson}.
In the present work, since an influence of this tiny FS may be small in the present analysis, we focus on the two main FSs observed by ARPES.
In this system, therefore, conductivity can be described in a nearly free electron model by keeping the effective mass ($m^*$) constant on each FS as follows:

\begin{equation}
    \sigma = \eta_1{\rm e}^2\tau_1 + \eta_2{\rm e}^2\tau_2.
\end{equation}
where $\eta_i = n_i/m^*_i$($i$=1,2) and $n_i$ is a carrier density of $i$-th FS.
From this equation we define a weighted average $\tau$ ($\tau_{\rm ave}$) as,
\begin{equation}
    \tau_{\rm ave} = \frac{\eta_1\tau_1 + \eta_2\tau_2}{\eta_1 + \eta_2}.
\end{equation}
Therefore,
\begin{equation}
    \sigma = (\eta_1 + \eta_2){\rm e}^2\tau_{\rm ave}.
    \label{sigma-tau}
\end{equation}
Thus, the relation between the residual resistivity ($\rho_0^{\rm AG} \equiv 1/\sigma$) and $\tau_{\rm ave}$, which should correspond to the $\tau$ in the Eq.\ref{AG_eq}, is derived.
The information on FeSe achieved from the ARPES measured at 30 K to calculate $\eta_i$ is shown in table I \cite{Nakayama,Nakayama_unp}.
Since the anisotropy of the hole FS at $\Gamma$ is smaller than the electron FS at M-point, the observed ARPES may represent such an averaged value.
For each FS, $m^*$ was calculated by the relation $m^* = \hbar^2k_{\rm F}/v_{\rm F}^{\rm ave}$.
The $v_{\rm F}$ was averaged over the electron FS and was evaluated as $v_{\rm F}^{\rm ave}$.
The general carrier density definition is,

\begin{equation}
    n = \frac{2}{(2\pi)^3}\int_{k\leq k_{\rm F}}{\rm d}k^3,
\end{equation}
where the factor of 2 is for the spin degeneracy and the integration is performed for the entire volume surrounded by FS.
Given the elliptic cylinder, $n$ can be calculated as,

\begin{equation}
    n = \frac{k_{\rm F}^{\rm l}k_{\rm F}^{\rm s}}{2\pi c},
\end{equation}
where $k_{\rm F}^{\rm l}$ and $k_{\rm F}^{\rm s}$ denote the Fermi wave number of the longer and the shorter axes of the ellipse, respectively.
The $c$-axis cell parameter expressed as $c$ in the real space is taken from literature \cite{Kosmas}.
Now we can calculate the $\rho_0^{\rm AG}$ in the framework of nearly free electron approximation from Eq. \ref{sigma-tau}.
The obtained parameters are $n_1 \approx 2.90\times10^{26}$ m$^{-3}$, $n_2 \approx 3.35\times10^{26}$ m$^{-3}$ and $m^*_1 \approx 2.28\times10^{-30}$ kg, $m^*_2 \approx 4.26\times10^{-30}$ kg, where the indices of 1 and 2 represent the electron and the hole bands, respectively.

\begin{figure}[h]
\includegraphics[width=0.8\linewidth]{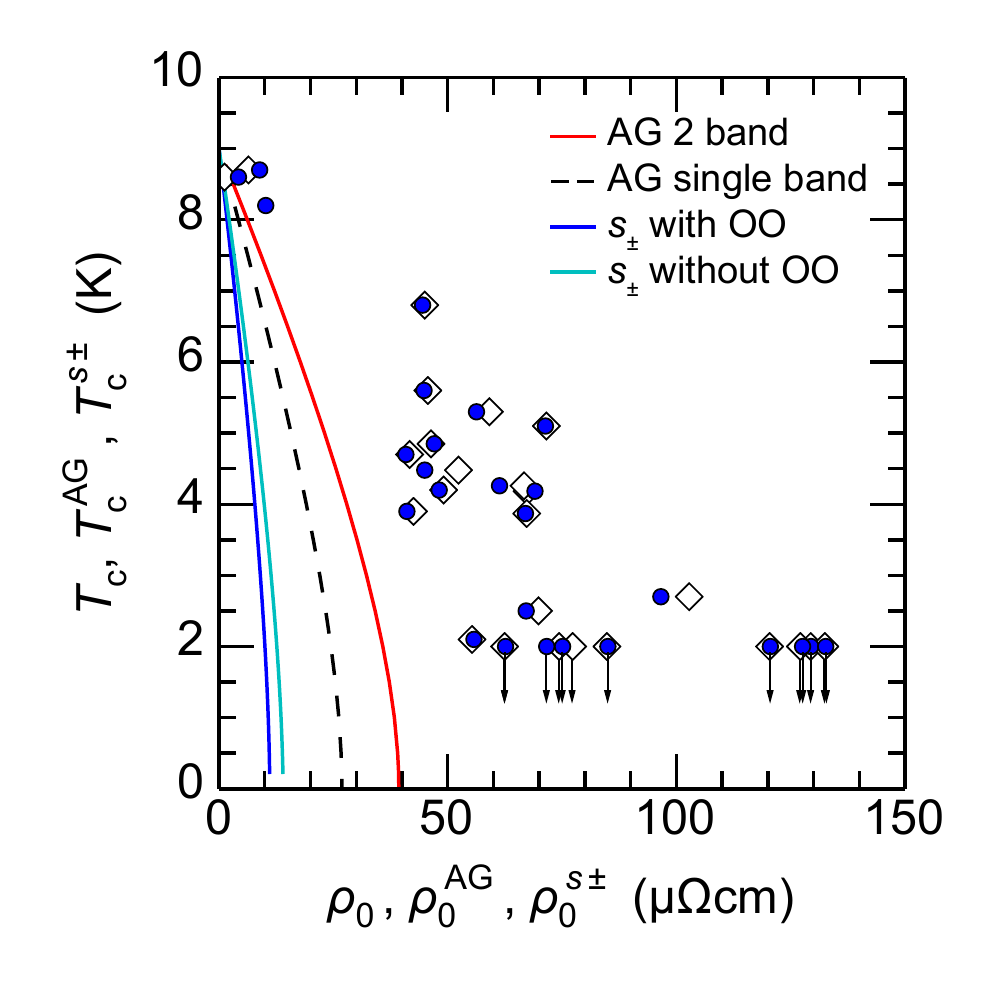}
\caption{
Dependence of the superconducting transition temperature ($T_{\rm c}$) on the residual resistivity ($\rho_0$).
The $\rho_0$ was estimated by fitting $\rho$-$T$ curves in the normal states as shown in Fig. \ref{res}(c).
White diamonds and blue circles represent $\rho_0$ estimated by $T$-linear and $T^\gamma$ type fitting functions, respectively.
With realistic band parameters, the red, blue and cyan lines are drawn by the AG-theory ($T_{\rm c}^{\rm AG}$ versus $\rho_0^{\rm AG}$) and the $s_{\pm}$-wave model ($T_{\rm c}^{s\pm}$ versus $\rho_0^{s\pm}$) with and without orbital order (OO), respectively.
The dashed line represents the AG-theory in the conventional single carrier model.}
\label{rr}
\end{figure}

The obtained dependence, i.e. $T_{\rm c}^{\rm AG}$ versus $\rho_0^{\rm AG}$, is represented by a solid red line in Fig. \ref{rr}.
The suppression of $T_{\rm c}^{\rm AG}$ is clearly faster than our experimental data.
Given a conventional single-band model, for comparison, the analytical expectation (dashed line) is represented  in the same figure, where the carrier number $n$ was evaluated from $R_{\rm H}$ data measured at low $B$'s.
The averaged effective mass was employed to calculate $\rho_0^{\rm AG}$.
The suppression of $T_{\rm c}^{\rm AG}$ is faster in the single-band model than that in the two-band model.
Since the amplitude of $R_{\rm H}$ is suppressed in a compensated semimetallic state, the single-band model ($R_{\rm H}$ = $1/n{\rm e}$) overestimates $n$ and consequently leads to faster suppression of $T_{\rm c}^{\rm AG}$.

\begin{table}[h]
\caption{Results of the ARPES measurements at 30 K \cite{Nakayama,Nakayama_unp}. Note that several photon energy results are averaged.}
\begin{tabular}{l c c c} \hline\hline
& $\Gamma$ & M (long axis) & M (short axis) \\ \hline
$k_{\rm F}$ (\AA$^{-1}$)& 0.11 & 0.20 & 0.05 \\
$v_{\rm F}$ (eV\AA) & 0.18 & 0.50 & 0.20 \\ \hline\hline
\end{tabular}
\end{table}

\subsubsection*{Electron doping effects on the two band model}
\begin{figure}[h]
\includegraphics[bb=0 0 407 191,width=1.0\linewidth]{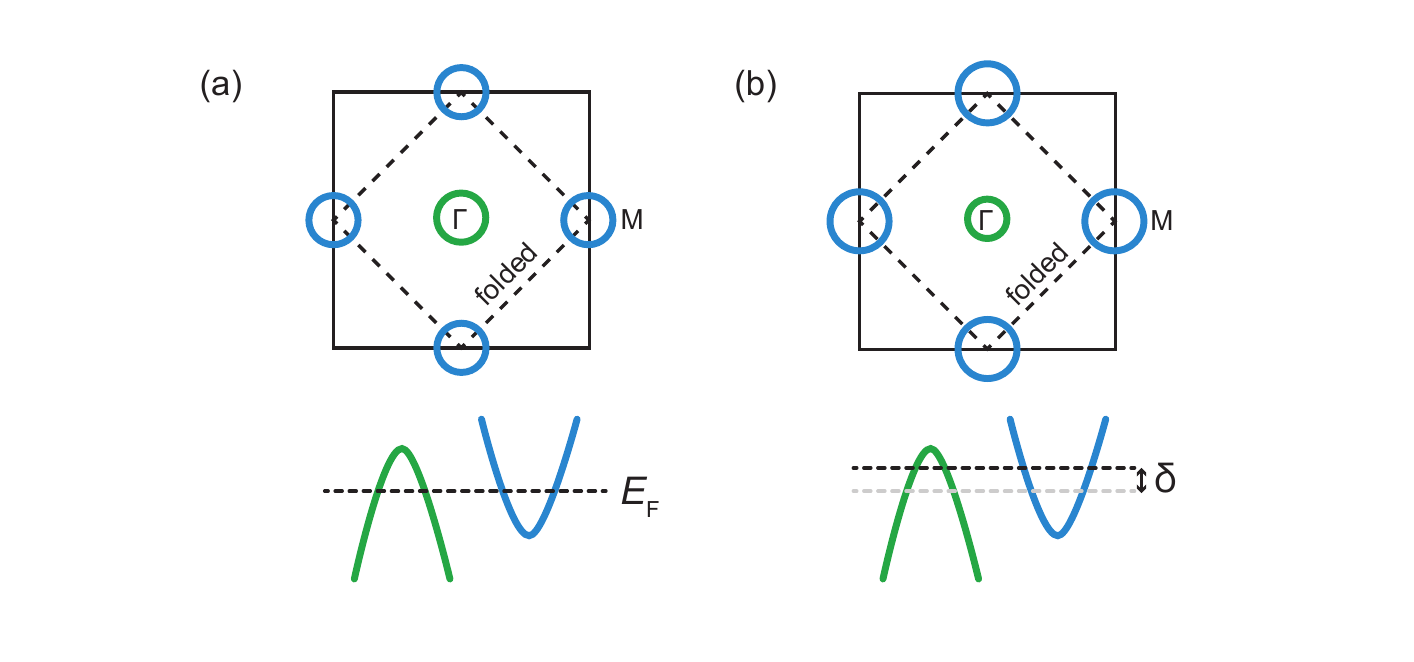}
\caption{A schematic view of Fermi surfaces of (a) two band semimetallic structure and (b) electron doped one on the $k_x-k_y$ plane. The bottom figure represents the rigid band Fermi energy shift due to the electron doping by $\delta$.}
\label{FS}
\end{figure}

We interpret the electron doping effects on our fitting model based on a nearly free electron model.
If the carrier doping rigidly shifts the Fermi energy ($E_{\rm F}$) as represented in Fig. \ref{FS}, the conductivity is shown not to be affected by $\eta_1+\eta_2$ but determined only by $\tau_{\rm ave}$ as follows:
In a nearly free electron model, energy dispersion relation of an electron band is written as,


\begin{equation}
E({\bf k}) = \frac{\hbar^2|{\bf k}|^2}{2m^*}-E_{\rm F}.
\end{equation}
The amplitude of Fermi wave number is derived by putting $E({\bf k})=0$ as $|{\bf k}_{\rm F}|=\pm\sqrt{2m^*E_{\rm F}}/\hbar$.
The carrier density of a 2 dimensional FS (a cylinder) is,

\begin{equation}
n = \frac{|{\bf k}_{\rm F}|^2}{2\pi c}=\frac{m^*E_{\rm F}}{\pi\hbar^2c}
\end{equation}
By assuming electron doping by $\delta$, $E_{\rm F}$ becomes $E_{\rm F}+\delta$.
The value of $m^*$ is not modified in a rigid band model.
Therefore the conductivity after the electron doping can be written as,


\begin{eqnarray}
\nonumber
\sigma &=& (\eta'_1 + \eta'_2){\rm e}^2\tau'_{\rm ave}\\
\nonumber
&=& \left(\frac{E_{\rm F}^{(1)}+\delta}{\pi\hbar^2c} + \frac{E_{\rm F}^{(2)}-\delta}{\pi\hbar^2c}\right){\rm e}^2\tau'_{\rm ave}\\
&=& (\eta_1 + \eta_2){\rm e}^2\tau'_{\rm ave}
\end{eqnarray}
where $E_{\rm F}^{(1)}$ and $E_{\rm F}^{(2)}$ represents the Fermi energy of electron and hole bands, respectively.
Therefore,  $(\eta_1 + \eta_2)$ of a 2 dimensional nearly free electron model does not change by rigid band shift.
As long as the approximation is valid, the effect of Co substitution can be simplified to the scattering time problem.

\subsection{Theory of impurity-induced reduction in $T_c$ for the full gap $s_\pm$-wave state}

\begin{figure}[htb]
\includegraphics[bb=0 0 595 842,width=0.9\linewidth]{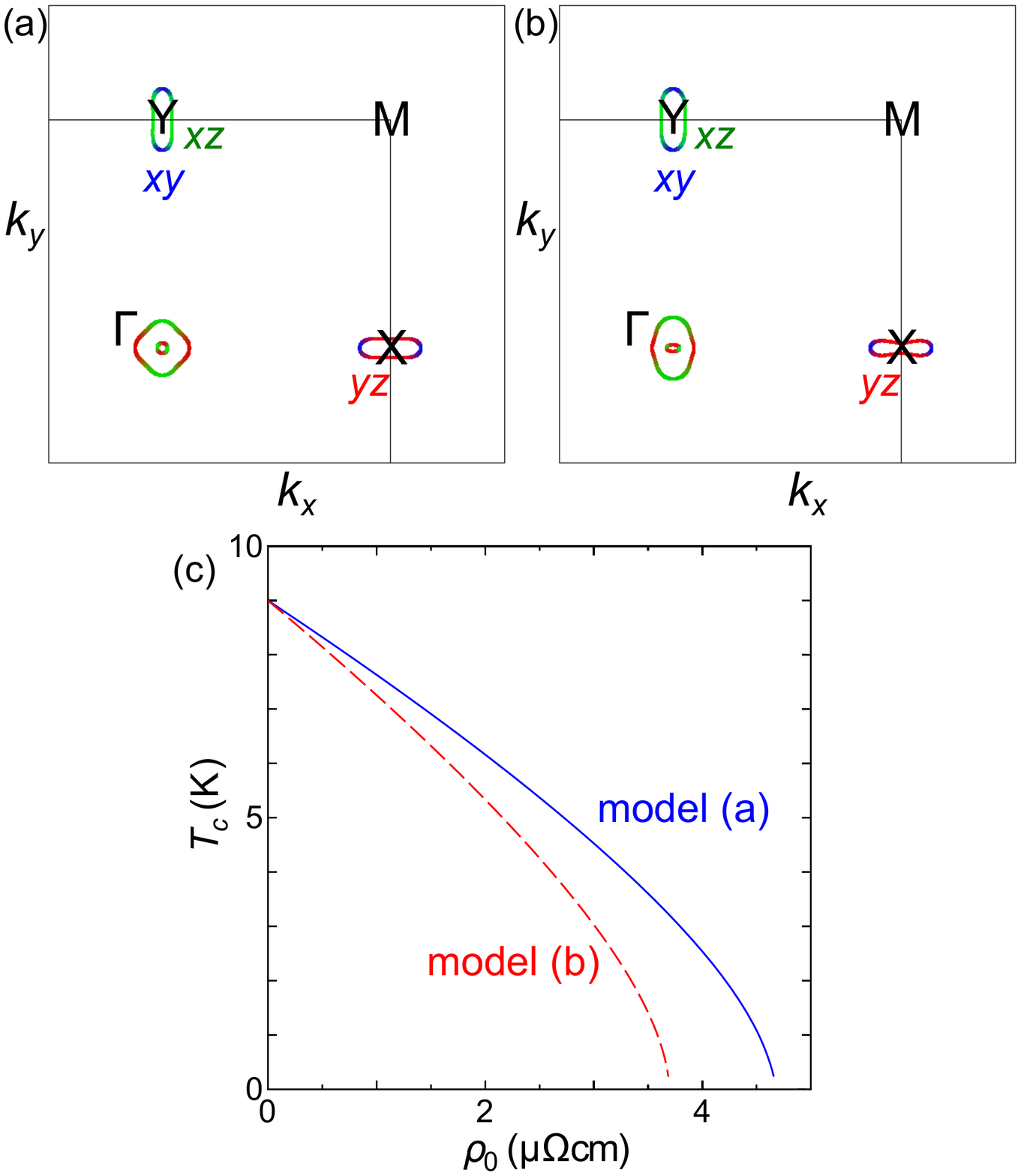}
\caption{(color online)
Fermi surfaces of the present tight-binding models for FeSe.
(a) without orbital polarization 
and 
(b) with orbital polarization with sign reversal
between $\Gamma$ and X(Y) points.
(c) Obtained impurity-induced reduction in $T_c$ for the $s_\pm$-wave state 
as a function of residual resistivity $\rho_0$ ($T_{\rm c}=9$ K), 
in both models (a) and (b).
}
\label{fig:theory}
\end{figure}

The second model to analyze the suppression of $T_{\rm c}$ is a realistic multiorbital theory.
Here, we employed a multiorbital tight-binding model for FeSe with realistic Fermi surfaces in order to perform quantitative theoretical studies of the impurity effect on the $s_{\pm}$-wave state.
Following two theoretical models are introduced:
In model (a), the orbital order is absent and the $C_4$ symmetry is preserved.
In model (b), the experimentally observed sign-reversing orbital polarization 
is introduced, by which the symmetry is lowered to $C_2$.
The Fermi surfaces of these models without carrier doping are shown in
Figs. \ref{fig:theory} (a) and (b), respectively.
The carrier numbers per Fe site are 
$n_h=2.02$\% and $n_e=2.02$\% in model (a), 
and $n_h=1.95$\% and $n_e=1.94$\% in model (b).
In model (a) (model (b)),
the hole-pockets disappear by introducing 
$3.0$\% ($3.7$\%) electron carriers induced by Co-doping.
This result would be consistent with the 
disappearance of the structure transition 
by $3.6$\% Co-doping as observed experimentally.

In order to construct the model (a) without the orbital order, 
we first performed LDA band calculations for FeSe
using the wien2k package, and made the tight-binding fitting
using the wien2wannier package.
However, the obtained Fermi pockets are too large.
To reproduce the tiny Fermi pockets experimentally observed in FeSe,
we introduced the orbital-dependent energy shifts at 
$\Gamma$, X(Y), and M points; $(\delta E_\Gamma, \delta E_X, \delta E_M)$:
We set 
$(\delta E_\Gamma, \delta E_{X,Y}, \delta E_M)=(-0.26,0.13,0)$ 
for $xz,yz$-orbitals, and  
$(\delta E_\Gamma, \delta E_{X,Y}, \delta E_M)=(0,0.26,-0.25)$
for $xy$-orbitals in the unit of eV.
Such level shifts were introduced by the additional 
intra-orbital hopping integrals;
$\delta t^{\rm on-site}=\delta E_\Gamma/4+ \delta E_{X,Y}/2+ \delta E_M/4$,
$\delta t^{\rm nn}=\delta E_\Gamma/8- \delta E_{X,Y}/8$, and
$\delta t^{\rm nnn}=\delta E_\Gamma/16+ \delta E_{X,Y}/16- \delta E_M/8$.
In the model (b),
we further introduced orbital polarization with sign-reversal:
$E_{yz}-E_{xz}=-0.02$ eV at $\Gamma$ and M points, and
$E_{yz}-E_{xz}=0.02$ eV at X and Y points.

Figure \ref{fig:theory} (c) shows the
impurity-induced reduction in $T_c$ of the $s_\pm$-wave state 
for $T_{\rm c0}=9$ K as a function of residual resistivity $\rho_0$,
obtained by applying the $T$-matrix approximation.
We used the realistic first principles impurity potential model for Co-ion given in \cite{Nakamura},
and neglected the carrier doping effect by Co doping.
Here, we introduced the effective intra-orbital and inter-pocket repulsive 
pairing interactions due to the spin fluctuations,
by adjusting the strength to realize $T_{\rm c0}=9$ K.
The detailed explanations for the gap equation analysis
had been given in \cite{Yamakawa}.
In the spin-fluctuation pairing mechanism in FeSe, 
only the electrons with $xz$ and $yz$ orbital characters contribute 
to the superconductivity since the $xy$-orbital is absent 
in the hole-pocket \cite{Mukherjee}.
We also put the renormalization for the $xz,yz$-orbitals ($z$) to be unity.
In both models (a) and (b), 
the superconducting state disappears when the 
residual resistivity is $\rho^{\rm cr}_0\approx 4 [{\rm \mu}\Omega{\rm cm}]$,
insensitively to the orbital polarization.
The current vertex correction for $\rho_0$ was taken into account,
and $\rho_0$ was kept independent of $z$.
If the experimental value $z\approx 1/3$ for the $xz,yz$-orbitals
is taken into account \cite{Maletz}, the critical residual resistivity is 
multiplied by $z^{-1}$.
Therefore, $\rho^{\rm cr}_0\approx z^{-1}4 [{\rm \mu}\Omega{\rm cm}]$.
Note that the total carrier number $n_{\rm carrier}$ is $\sim$4\% per Fe ion in both model (a) and model (b).
In a model with $n_{\rm carrier}$ $\sim$2.5\% by reducing the sizes of Fermi pockets,
the relation $\rho^{cr}_0 \approx z^{-1}6.5[{\rm \mu}\Omega{\rm cm}]$ is realized.

Based on the $\rho^{\rm cr}_0\approx z^{-1}4 [{\rm \mu}\Omega{\rm cm}]$ with $z^{-1}$ $\sim$3, the superconducting transition temperature ($T_{\rm c}^{s\pm}$) suppression predicted in the $s_{\pm}$-wave state is displayed by the blue (the orbital order is considered) and cyan (the orbital order is not considered) lines in Fig. \ref{rr}.
In both cases, the calculated results show faster reductions of $T_{\rm c}^{s\pm}$ than the experimental data, being consistent with that in the first model.

\begin{figure}
\includegraphics[width=1.0\linewidth]{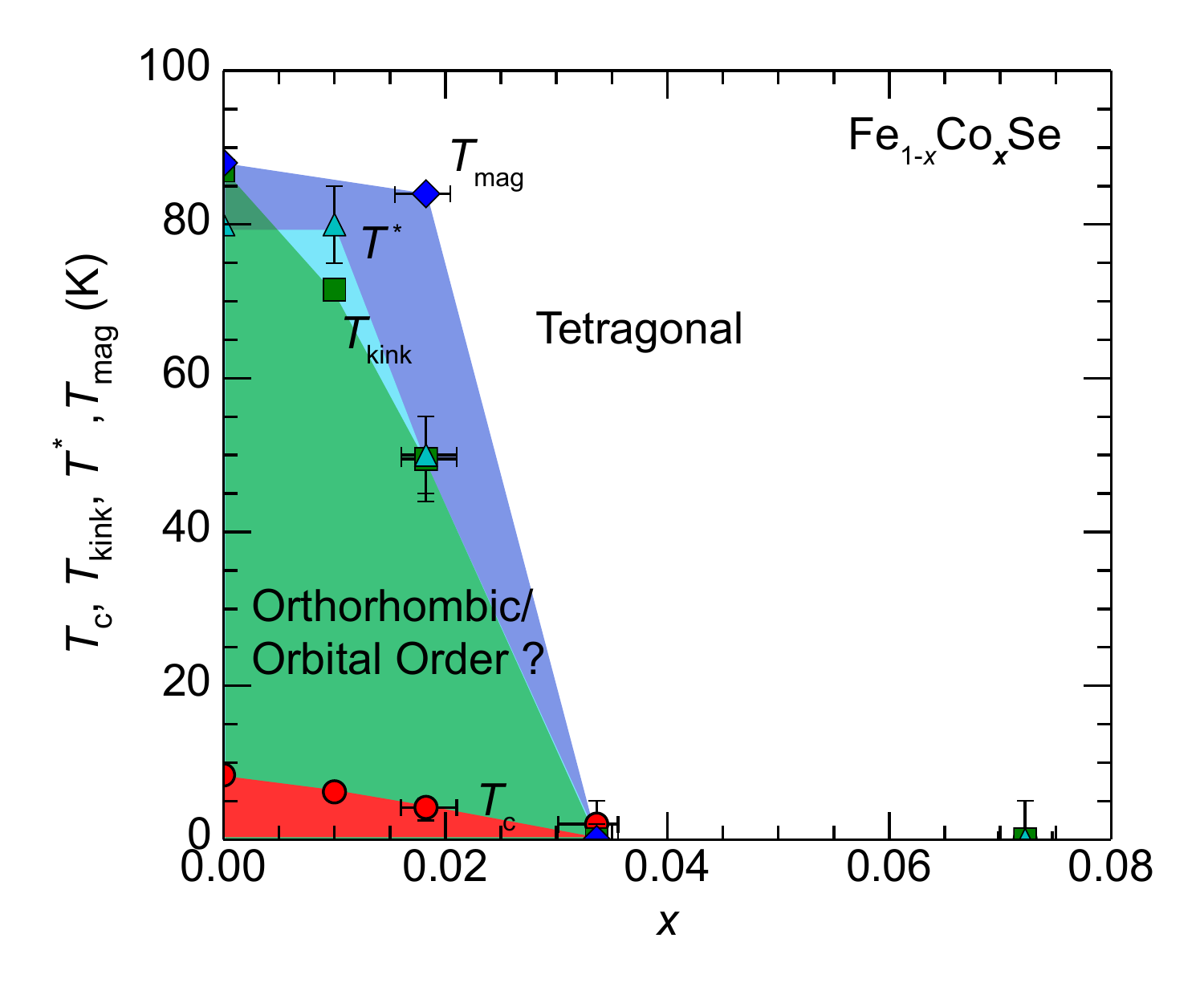}
\caption{An electronic phase diagram of Fe$_{1-x}$Co$_x$Se. The superconducting transition temperature ($T_{\rm c}$ : red circles), the temperature showing a kink in the $\rho$-$T$ curve ($T_{\rm kink}$: green squares), and temperature showing a sudden decrease in the Hall coefficient ($T^*$: light blue triangles) are represented.
$T_{\rm mag}$ (blue diamonds) represents a kink observed in temperature dependence of magnetic susceptibility.}
\label{phase}
\end{figure}

\section{Discussion}
\subsection{Electronic Phase diagram}
The experimental observations in Fe$_{1-x}$Co$_x$Se are summarized in a phase diagram shown in Fig. \ref{phase}.
The $R_{\rm H}$ at low temperatures increases first and then decreases with an increase in $x$ of Co as shown in Fig. \ref{Hall_coeff}.
Above $x$ = 0.036, $R_{\rm H}$ becomes nearly constant against $x$.
In a two band semi-classical model, the low field limit of Hall coefficient can be written as,

\begin{equation}
    R_{\rm H} = \frac{p\nu^2 - n\mu^2}{e(p\nu+n\mu)^2},
    \label{Hall_co}
\end{equation}
where $n$, $p$ and $\mu$, $\nu$ are the carrier concentrations and the mobilities of holes and electrons, respectively.
ARPES, STM and transport studies have demonstrated that the electronic structure in the orthorhombic FeSe is almost compensated semimetal \cite{Nakayama,Shimojima_FeSe,Kasahara_PNAS,Huynh}.
In a compensated semimetal ($n\approx p$), the amplitude of $R_{\rm H}$ is suppressed due to the electron and hole compensation as shown in eq. \ref{Hall_co}.
When electrons are doped to FeSe, the compensation breaks to increase the amplitude of $R_{\rm H}$.
Further electron doping may shrink and vanish the tiny hole pocket.
In such a case, since a single carrier picture is more appropriate, the amplitude of $R_{\rm H}$ is suppressed.
Therefore, our experimental observations on the Co doping dependence  $R_{\rm H}$ in Fe$_{1-x}$Co$_x$Se can qualitatively be understood in terms of the electron doping by Co.
For $x$ = 0.036 and 0.075, where the high temperature tetragonal phase remains in an entire temperature range, the $R_{\rm H}$ at 10 K gives the carrier number ($n$) of $\sim$ 10$^{21}$ cm$^{-3}$ in a single carrier model (i.e. $R_{\rm H}$ = 1/$n$).
By assuming the rigid band shift of the Fermi energy in Fe$_{1-x}$Co$_x$Se, the carrier number in the electron pocket above $x$ = 0.036 is deduced to be $\sim$ 10$^{\rm 21}$ cm$^{-3}$ based on the carrier numbers of the electron and the hole pockets in FeSe \cite{Huynh}, being comparable with those in $x$ = 0.036 and 0.075.
Since the variation observed in $R_{\rm H}$ is consistent with the electron doping provided by Co, potential scattering and changes in both the type and the number of carriers influence on the phase diagram in Fig. \ref{phase}.
Note that quantitative analysis centering on the Hall resistivity curvature is not realistic in the present magnetic field regime.
The transverse magnetoresistance is $\approx 10$\% at $T = $10 K and $B = $ 9 T for $x \approx 0.018$ whereas it is $\approx 300$\% in the parent FeSe \cite{Huynh}, suggesting the significant suppression in carrier mobility by the Co scattering effects.
Since only a limited number of carriers contribute to the magnetotransport below 9 T, any reasonable analysis on the magnetotransport data would not accurately be valid \cite{Kim_SM,Urata_Mn_mobility}.
Semiclassical multicarrier fitting of Hall resistivity assuming the number of electron and hole pockets could not give correct transport parameters for anisotropic Fermi surfaces, and somewhat more sophisticated analyses such as a mobility spectrum method are necessary \cite{Huynh}.
For preventing these uncertain factors in discussion, we employed the band parameters extracted from the ARPES data for the present discussion.

Both superconductivity and structural transition are suppressed by Co doping and disappear at $x$ = 0.036.
This is consistent with the disappearance of the kink in magnetic susceptibility curves ($T_{\rm mag}$).
Since the orbital order was reported to develop in the orthorhombic phase of FeSe, the present results may also indicate the suppression to be correlated between superconductivity and orbital order by Co doping.
Theoretically, nesting between electron and hole pockets is predicted to be important for both superconductivity and orbital order \cite{Mazin,Kuroki,Kontani} in FeSCs \cite{Chubkov,Hosono_Kuroki,Onari}.
In FeSe, electron doping may also suppress the nesting due to the small Fermi energies \cite{Kasahara_PNAS}, being consistent with the experimentally observed evolution of the $R_{\rm H}$ as a function of $x$.
Hence, the suppression correlated among superconductivity, structural phase transition, and orbital order may originate from the reduction in nesting between electron and hole pockets.

\subsection{Paring Mechanism}
If the sign reversal superconducting mechanism could be dominant in FeSe, suppression in superconductivity would be similar to that predicted by both theories of AG and the full gap $s_{\pm}$-wave states.
However, the suppressions experimentally observed as $T_{\rm c}$ was much smaller than those described in these models.
Consequently, the sign preserved superconducting mechanism will be more appropriate for Fe$_{1-x}$Co$_x$Se.
In FeSe, a nodal superconducting gap was experimentally observed in the electron pocket \cite{Kasahara_PNAS,Watashige}.
Meanwhile, orbital resolved ARPES demonstrated an imperfect antiferromagnetic nesting between the electron and the hole Fermi surfaces.
One can imagine that the sign reversal and the sign preserved pairing mechanism may compete with each other via the inter and the intra orbital scattering \cite{Mazin,Kuroki,Kontani,Onari,Hosono_Kuroki}.
In this case, the sign preserved pairing mechanism may give a primary contribution in FeSe from the viewpoint of the $T_{\rm c}$ suppression observed in our experimental data.
Another possible scenario may be that the superconducting gap structure changes from a nodal $s$- to an $s_{\rm ++}$-wave states.
It has been pointed out that the node of the superconducting gap is not protected by symmetry, and an accidental node \cite{Kasahara_PNAS,Dong} was suggested.
The superconducting gap may change from a nodal gap to a full gap state depending on the quality of single crystals.
In such a case, the potential scattering induced by Co may lift the node to open a superconducting gap, resulting in a change of gap structure from the nodal- to the $s_{\rm ++}$-wave states.
It should be noted that sign preserved superconducting gap structure has been proposed for single-layered FeSe on SrTiO$_3$ \cite{Q_Fan} possessing highly electron doped band structure \cite{D_Liu}.
Our observations are consistent with these previous works.

\section{Conclusion}
We investigated the nonmagnetic Co impurity doping effect on the superconductivity and the structural transition/orbital order in FeSe.
In the electronic phase diagram of Fe$_{1-x}$Co$_x$Se, both superconductivity and structural transition/orbital order were suppressed by Co doping and disappeared at $x$ = 0.036.
Experimental data of the $T_{\rm c}$ suppression against the residual resistivity were compared to the values expected from the AG theory \cite{AG} and a full gap $s_{\pm}$ theory \cite{Yamakawa}.
Our data showed much slower suppression than those from both theories.
As a result, the sign preserved superconducting mechanism is more appropriate to explain the superconducting pairing mechanism in FeSe.
In FeSe, the interband Fermi surface nesting including both intra- and inter-orbital scattering processes play an essential role for the superconducting pairing mechanism and for the normal state phase diagram \cite{Chubkov,Onari,Hosono_Kuroki}.
The present studies unveiled a part of the complex electronic states.
Considering the electronic structures taking place in this system, small perturbation may change the situation more complex in the Fe-3$d$ multiband system by changing one state to another.

\section{Acknowledgements}
The authors are grateful to K. Nakayama for providing the unpublished ARPES results of Fe$_{1-x}$Co$_x$Se and fruitful discussions. 
One of the authors (T.U.) was supported by the Research Fellowship of Japan Society for the Promotion of Science.

\bibliography{FeCoSe}
\end{document}